# Sequential Circuits Synthesis for Rapid Single Flux Quantum Logic Based on Finite State Machine Decomposition

Shucheng Yang, Xiaoping Gao, Jie Ren

*Abstract*—Rapid Single Flux Quantum (RSFQ) logic is a promising technology to supersede Complementary metal-oxide-semiconductor (CMOS) logic in some specialized areas due to providing ultra-fast and energy-efficient circuits. To realize a large-scale integration design, electronic design automation (EDA) tools specialized for RSFQ logic are required due to the divergences in logic type, timing constraints, and circuit structure compared with CMOS logic. Logic synthesis is crucial in converting behavioral circuit description into circuit netlist, typically combining combinational and sequential circuit synthesis. For the RSFQ logic, the sequential circuit synthesis is challenging, especially for non-linear sequential blocks with feedback loops. Thus, this paper presents a sequential circuit synthesis algorithm based on finite state machine (FSM) decomposition, which ensures design functionality, lowers costs, and improves the RSFQ circuit performance. Additionally, we present the synthesis processes of the feedback logic and the 2-bit counter to demonstrate how the proposed algorithm operates, and ISCAS89 benchmark circuits reveal our method's ability to synthesize large-scale sequential circuits.

## I. INTRODUCTION

Rapid Single Flux Quantum (RSFQ) logic [1] is a novel digital logic that uses the superconducting process during manufacturing. The fundamental element in the RSFQ logic is the Josephson junction (JJ) that transmits an Single Flux Quantum (SFQ) signal at a picosecond level and dissipates around $10^{-19}$ J per switching, contributing to ultra-fast and energy-efficient circuits and systems.

To design a large-scale RSFQ circuit, electronic design automation tools (EDA) are required. However, the differences between the RSFQ and Complementary metal-oxide-semiconductor (CMOS) gates raise the RSFQ digital design automation complexity. Specifically, in the RSFQ gates, the basic Boolean gates, such as AND, OR, and NOT are clocked, as the Boolean value "True" and "False" in the RSFQ logic are represented by the existence of an SFQ pulse during a specific time window. This imposes various changes in several digital design automation steps, especially logic synthesis. Generally, the RSFQ logic synthesis comprises two categories: combinational and sequential circuit synthesis. The former synthesis has already been considered by many researchers [2-5], most of whom exploited existing Boolean logic synthesizers such as yosys [6] and ABC [7], which were appropriately modified to make them compatible with the RSFQ logic. Nevertheless, sequential circuit synthesis for RSFQ logic is hard to complete owing to its clocked Boolean gates. This paper proposes a sequential circuit synthesis algorithm for RSFQ logic based on finite state machine (FSM) decomposition. Our algorithm solves the sequential circuit synthesis problem and improves the circuit's performance including clock frequency, area and power consumption.

The remainder of this paper is organized as follows. Section II introduces the problems in RSFQ sequential synthesis and presents some RSFQ gate modeling examples. Section III introduces the proposed sequential circuit structure and the related algorithms, while Section IV performs small-scale case studies and a large-scale experiment. Finally, Section V concludes this work.

## II. SYNTHESIS ISSUES FOR RSFQ LOGIC

### A. Previous Works on RSFQ Logic Synthesis

For RSFQ logic synthesis, most works focused on optimizing the combinational circuits. As mentioned above, the RSFQ circuits are naturally pipelined since the gates are

This work was supported by the Strategic Priority Research Program of Chinese Academy of Sciences (Grant No. XDA18000000), Shanghai Science and Technology Committee (Grant No. 21DZ1101000), the National Natural Science Foundation of China (Grant No. 62171437 and 92164101). (Corresponding author: Jie Ren).

Shucheng Yang and Xiaoping Gao are with the Shanghai Institute of Microsystem and Information Technology, Chinese Academy of Science, Shanghai 200050, China and also with the CAS Center for Excellence in Superconducting Electronics (CENSE), Shanghai 200050, China.

Jie Ren is with the Shanghai Institute of Microsystem and Information Technology (SIMIT), Chinese Academy of Science (CAS), Shanghai 200050, China, with the CAS Center for Excellence in Superconducting Electronics (CENSE), Shanghai 200050, China and also with the University of Chinese Academy of Science Beijing, Beijing 100049, China (e-mail: jieren@mail.sim.ac.cn).

clocked. Thus, several techniques are required to ensure the circuit's functionalities and performance, such as path balancing [2] and supergates [8]. Path balancing is a technique for balancing the difference in logical depth between logic gates in a Boolean logic netlist with a chain of D flip-flops, making it an RSFQ netlist. Supergates are small-scale circuits designed for specific functions and can be used for technology mapping. A typical synthesis process of the RSFQ combinational circuit is illustrated in Fig.1.

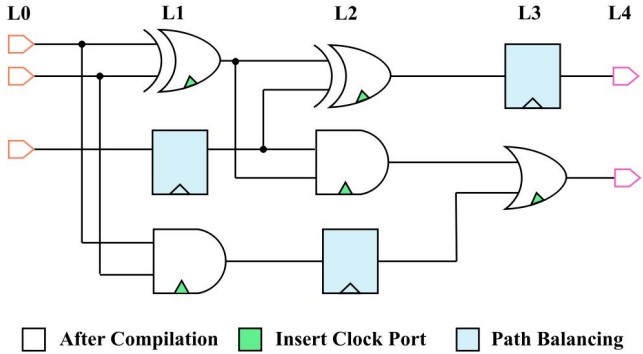

**Fig.1.** Post-treatment for RSFQ combinational circuits after standard compilation. Non-clock gates must be converted into clocked by inserting clock ports, and D flip-flops must be added to balance the path. L0 to L4 are the logic depth columns.

By employing the same parser and optimizer as CMOS logic, we obtain the original Boolean logic netlist after compilation. However, this compilation result cannot be used directly as an RSFQ netlist since the original Boolean logic gates are non-clocked and the path is not balanced in terms of logic depth. Therefore, we must convert the non-clocked gate reference into clocked gate reference, i.e., insert a clock port at each logic gate, e.g., AND, OR, and NOT. To balance the circuit's path, the logic depth of each gate in terms of inputs is calculated, and the D flip-flop chain, with its length derived according to the logic depth difference, is inserted into the netlist. For example, the logic depth difference between the RSFQ-AND gate in L1 and the RSFQ-OR gate in L3 is 1, so one D flip-flop is inserted between these two gates. Moreover, the fanout of each net is optimized by inserting splitter trees.

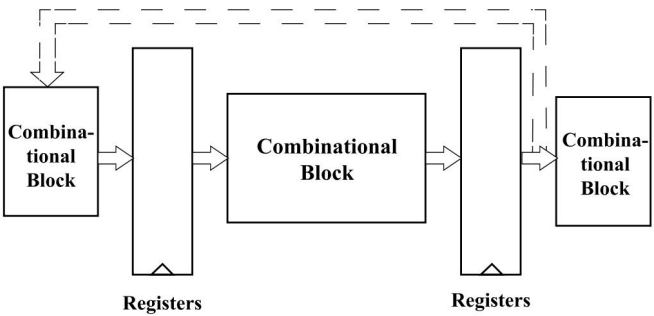

**Fig.2.** Sequential logic structure. Neglecting the feedback loop converts the structure into a simple pipeline.

Regarding the sequential logic, two sequential circuit types exist. The first is a pipeline structure with no loop, typically treated as the combinational logic with similar path balancing techniques. The second type is more complex, comprising sequential and combinational blocks with a feedback loop (Fig.2). To our knowledge, only one paper fully demonstrates how to synthesize a sequential RSFQ circuit [9].

### B. Problems in RSFQ Sequential Synthesis

CMOS logic synthesis tools present the sequential target circuit from the input design description. However, since the RSFQ gate is self-clocked, the timing sequence may be wrong if nothing is applied to the synthesis result. In [9], the authors presented a sequential synthesis method for RSFQ logic by introducing the path balancing technique for sequential synthesis. Nevertheless, many flip-flops are required, thus increasing the circuit area and the power consumption. Furthermore, although this method provides a functional sequential RSFQ circuit, the output frequency would be divided in terms of the loop's length. For example, the 2-bit counter in [9] reveals that the output is driven by every five clock inputs since the loop length regarding the latches is five. The inconsistency induced by changing the frequency occurs when this 2-bit counter is connected to other blocks. Specifically, such dysfunctionalities are caused by the mismatch between the output frequency of the former block and the input sample rate of the next block.

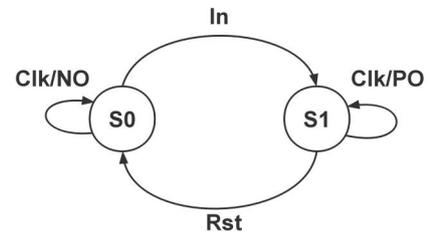

**Fig.3.** The State-Transition Diagram of the FSM example.

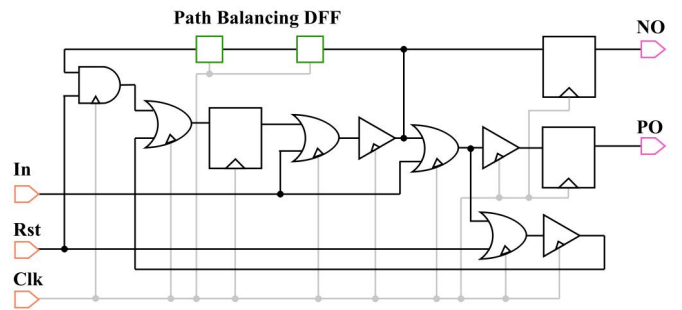

**Fig.4.** Netlist derived from the logic synthesis tool for CMOS logic, with clock ports and path balancing inserted.

Under some extreme circumstances, the target functionalities may not be guaranteed. For instance, the FSM illustrated in Fig.3 is synthesized by a logic synthesis tool appropriate for CMOS logic, which is then processed using the path balancing technique. The output netlist is depicted in

Fig.4. However, the simulation results in Fig.5 highlight that even though the positive output (PO) waveform is correct and its output frequency is divided, the negative output (NO) fails to match its designed functionality since there should be no SFQ pulses at the NO after the state transition to S1. This is because the state signal frequency is also divided by the loop's length, resulting in the FSM system only staying at S1 every seven clock inputs after the state transition to S1. In conclusion, the traditional synthesis method limits the possibility of synthesizing an RSFQ sequential circuit.

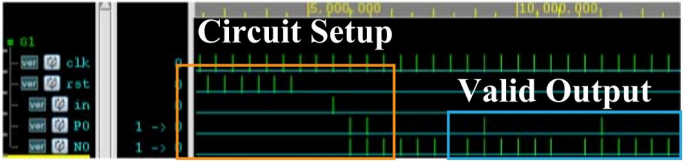

**Fig.5.** Simulation waveform of the above netlist. Circuit Setup is the intermediate process of circuit operating after changing the input signals, and Valid Output means the output data of PO and NO are valid only after the completion of Circuit Setup. Note that NO is valid after Circuit Setup but is incorrect.

*C. RSFQ Gates Behavioral Modeling*

RSFQ gates are usually modeled using a finite state machine [10-11], such as D Flip-Flop (DFF), Non-Destructive Read Out (NDRO), Resettable Toggle Flip-Flop (RTFF), and Complementary D Flip-Flop (DFFC) (Fig. 6). These 2-state gates have the simplest form and are indecomposable. It is worth noting that the RSFQ Boolean gates, such as AND, could also be modeled using FSM since the equivalent form of these gates comprise registers and the corresponding Boolean operation (Fig. 7).

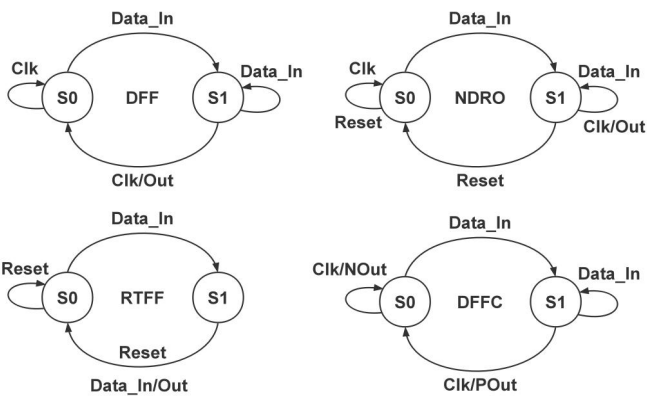

**Fig.6.** FSM models for DFF, NDRO, RTFF, and DFFC cells.

Since almost all RSFQ gates and sequential circuits can be described using FSM, it would be natural to consider that using small FSM of RSFQ gates to directly compose a large-scale sequential circuit is feasible. In a system design, this process is usually reversed, with a more practical way involving decomposing a large-scale FSM into small ones and mapping the small FSMs into RSFQ gates. Next, we will discuss the detailed techniques.

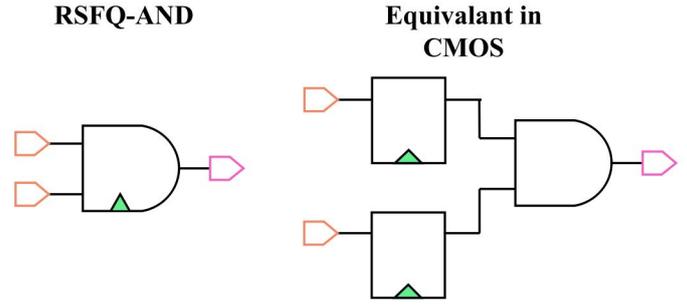

**Fig.7.** RSFQ AND gate and its equivalent Boolean logic circuit in CMOS logic.

III. SEQUENTIAL CIRCUIT STRUCTURE AND ALGORITHM

*A. Sequential Circuit Structure for RSFQ logic*

To synthesize the RSFQ sequential circuit, we divide the sequential RSFQ system into two parts, namely a combinational transition signal generation (CTSG) and state transition (ST) (Fig.8). The CTSG part is used for generating the input transition signals from the FSM extraction results. In some complex FSM, the input transition signal may be Boolean logical. Thus, a combinational block is needed. Meanwhile, in simple FSMs, these blocks could be just a set of input signal wires. For the RSFQ logic, this CTSG block can be optimized and balanced using the existing techniques mentioned in Section II-A. Furthermore, the state transition part is used to accomplish the states' transition and is the core block in the sequential logic. In this article, we focus on the ST part.

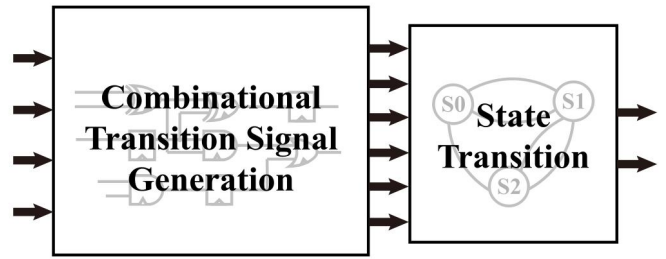

**Fig.8.** Generalized sequential circuit structure for RSFQ logic. Combinational transition signal generation is needed only when the input transition signal is a complex expression.

The overall synthesis flow is summarized in Fig.9. Note that the combinational blocks are not included in the sequential synthesis flow even if they exist in the Register-Transfer Level (RTL) design, as they will be synthesized using the methods of Section II-A. To speed up the synthesis process, parallel computation is applied to the state decomposition and sub-FSM mapping for each state encoding.

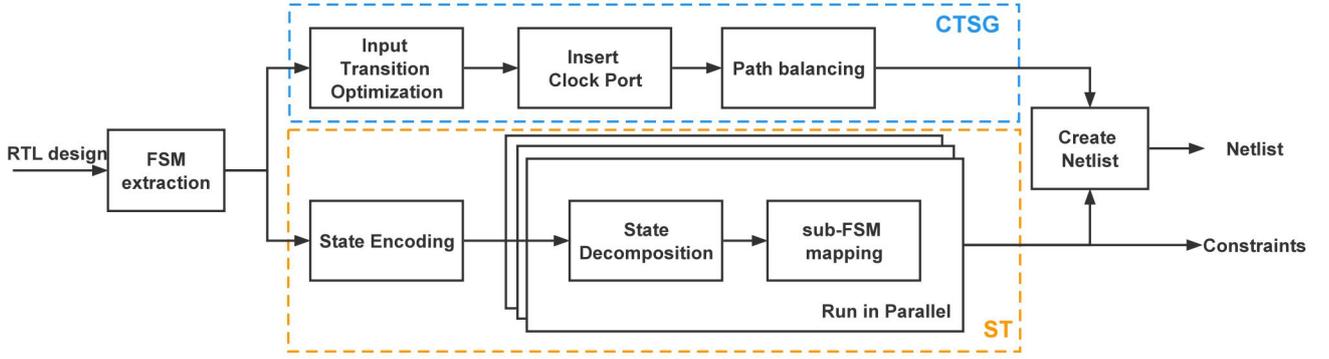

**Fig.9.** Flow chart of synthesizing RSFQ sequential circuits.

Once there is a valid technology mapping for one thread, the procedure is terminated, and the final results are output. Regarding the state decomposition itself, using the parallel algorithm, such as optimizing the input transition table in parallel, may be inefficient since the number of variables in the input transition table to be optimized is usually small. Thus, only serial computation is applied for state decomposition.

*B. Finite-State-Machine Extraction*

FSM extraction is a relatively mature technique, widely used in commercial and open-source tools. We choose the open-source Verilog parser [12] for FSM extraction, with the extraction tool solely designed for the RTL description, i.e., only the source code of the FSMs' description is acceptable. Although studies on extracting FSM from a circuit netlist [13-14] are available, these will not be discussed here since it is not the main topic of this article.

*C. States In RSFQ Perspective*

For the RSFQ logic, the input signal leads to several known results: (1) Stores a flux quantum in the loop. (2) Clears the flux quantum inside the loop. (3) Toggles the loop's state. (4) Outputs a flux quantum if a flux quantum is inside the loop. (5) Outputs a flux quantum if no flux quantum is inside the loop. For example, in DFFC cells, the input Data_In is (1), clk is (2), (4), and (5); In RTFF cells, the input Data_In is (3), and Reset is (2), according to the FSM model given in Fig.6. This background knowledge is summarized in Table. I and will be used in the state decomposition.

Table I. Input transition and Output Types

|  | Expression | Type |
| --- | --- | --- |
| SFQ set | $Q + F$ | Transition |
| SFQ clear | $Q \cdot \overline{F}$ | Transition |
| Toggle State | $Q \oplus F$ | Transition |
| Clock Out | $Q \cdot F$ | Output |
| Clock Negative Out | $\overline{Q} \cdot F$ | Output |

Note that the expression could be complex based on the abovementioned possibilities since one input transition signal may have multiple effects in one gate. For example, $Q \cdot \overline{F_1} + F_2$ means $F_1$ is the SFQ clear while $F_2$ is the SFQ set, where $Q$ is one of the state's sequential components.

*D. State Encoding*

State encoding is essential for the RSFQ sequential logic decomposition since it determines the number of processes used in the parallel optimization part. For the real RSFQ circuit, if the flux quantum is stored in the loop, the current state is TRUE (or 1). However, we do not know the actual circuit structure before we obtain the final synthesis result. If we search for all possible state encodings, the number of iterations will be n! for an FSM containing n states, which is very large when considering large-scale FSM. Therefore, we employ a heuristic method to sort the possible state encodings and create a state encoding tree to solve this problem. Fig.10 illustrates an example of an FSM and its state encoding tree. After deriving the state encoding tree, the parallel optimization procedure creates the state decomposition tasks, starting from the leftmost to the rightmost path of the tree.

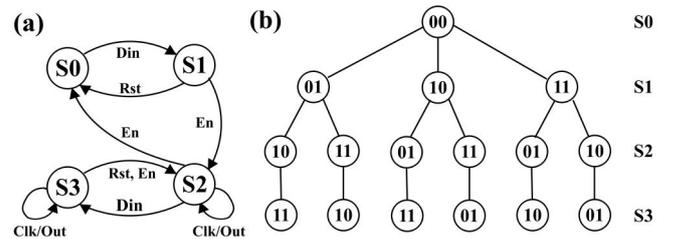

**Fig.10.** (a) Example of the FSM and (b) its state encoding tree. The FSM describes an enabled-resettable D-flip flop (ERDFF) logic.

First, the initial state should be set to all-zero since, at the real initial state of the RSFQ chip, as there are no SFQs across the whole circuit. Ideally, the input transition signal should affect only one sequential component, in which case the synthesis result will be the simplest. So, we set the state encoding by the input transition signal individually, then try the state encoding where one signal may affect two sequential components and keep this process until all the states are exhausted. The possible state encoding is accomplished between two states stepwise, and if the input signal is causing a state transition, it may be the the set or toggle. Accordingly, an input signal that makes the state return to the original state can be an SFQ clear. Once the possible type of the input signal is selected, it should be recorded and used in the following iterations. The algorithm is described as follows.

**Algorithm 1:** State Encoding Tree

1: **Input**: FSM
2: **Output**: Possible State Encoding Tree
3: Set Initial State S0 as all-zero (000...0)
4: StateTree←initialize()
5: StateTree.addNode(S0)
6: nodes←DFS(FSM,start←S0)
7: checkNode←set()
8: checkNode.add(S0)
9: sigTypeDict←initialize()
10: **while** True **do**
11:   **if** checkNode **is** empty **then**
12:     **break**
13:   currState←checkNode.pop()
14:   **for** nextState **in** nodes[currState] **do**
15:     checkNode.add(nextState)
16:     sigList←FSM[currState, nextState]
17:     **for** sig **in** sigList **do**
18:       **if** sig **not in** sigTypeDict **then**
19:         sigType ← guess the signal type based on the current state, next state, and the transition
20:         relationship.
21:       sigTypeDict.update(sig:sigType)
22:     possibleCodes←generate possible codes according to states and signal type
23:     StateTree.addNodes(possibleCodes)
24: **end**

### E. State Decomposition

There are various methods given in [15-19] to decompose large FSM systems, which are primarily graph-based and thus hard to understand and utilize in RSFQ logic. From the RSFQ logic perspective, each state bit is a sequential component, i.e., a 3-bit state can be decomposed into three sequential components, representing three sequential components. Thus, we decompose the large FSM simply into sequential components based on the bit number. An example is presented next utilizing the FSM of Fig.10 and its state encoding in Table II.

Table II. Example State Encoding

| S0 | S1 | S2 | S3 |
|---|---|---|---|
| 00 | 01 | 10 | 11 |

The input transition table of the given FSM is summarized in Table.III. The input transition signals are independent of the RSFQ logic since the state of an RSFQ system changes instantly once the input signal arrives. Thus, the state table optimization can be done individually. Additionally, the Boolean logic optimization can be done using open-source Binary Decision Diagram (BDD) optimizers, such as [20]. The optimized expressions of each input signal and state component could be used to mark the input signal type.

Table III. Input Transition Table

| State | | Din | | Rst | | En | | Clk | |
|---|---|---|---|---|---|---|---|---|---|
| Q1 | Q0 | 0 | 1 | 0 | 1 | 0 | 1 | 0 | 1 |
| 0 | 0 | 00 | 01 | 00 | 00 | 00 | 00 | 00 | 00 |
| 0 | 1 | 01 | 01 | 01 | 00 | 01 | 10 | 01 | 01 |
| 1 | 0 | 10 | 11 | 10 | 10 | 10 | 00 | 10 | 10/1 |
| 1 | 1 | 11 | 11 | 11 | 10 | 11 | 10 | 11 | 11/1 |

The optimized expressions of each input transition signal are summarized in Table.IV and the input transition signal type are marked based on Section III-C. Simple optimized expressions, such as $Q0^* = Q0 + Din$, could be directly mapped according to Table.I. Meanwhile, the primary state variable should be extracted for complex expressions such as $Q1^* = Q1 \cdot \overline{En} + Q0 \cdot En$, and the optimized result should be reformulated. For example, in Table.IV, $QN$ stands for each bit of the current state while $QN^*$ stands for the next state, and the optimized expression $QN^* = QN$ is the holding state, i.e., this sequential component is irrelevant to the input, it won't be changed when the input arrives. The relationships between $Q0$ and $Din$, $Q0$ and $Rst$, $Q0$ and $En$ are presented in Table.I. The complex expression of $Q1$ and $En$ are of two parts, $Q1 \cdot \overline{En}$ means $En$ is an SFQ clear for $Q1$, and $Q0 \cdot En$ is an output expression of $Q0$ and an SFQ set for $Q1$, so the nets are marked for the following mapping process. After all types of input signals for each sequential component are marked based on the above process, the sub-FSM mapping procedure will be activated.

Table IV. Decomposition Results

| | Q1 | Q0 |
|---|---|---|
| Din | $Q1^* = Q1$ | $Q0^* = Q0 + Din$ |
| Rst | $Q1^* = Q1$ | $Q0^* = Q0 \cdot \overline{Rst}$ |
| En | $Q1^* = Q1 \cdot \overline{En} + Q0 \cdot En$ | $Q0^* = Q0 \cdot \overline{En}$ |
| Clk | $Q1^* = Q1$ | $Q0^* = Q0$ |
| Output | $Out = Q1 \cdot Clk$ | |

## F. Sub-FSM mapping

The technology mapping process can be done in terms of the markers. The target cells in the library are encoded based on the input port type, and the encoded tuples are sorted. For example, the RDFFC gate in the library is encoded as (0,1,6), which is a unique ID. Thus, a hash table can be applied for the mapping procedure. Once there is a decomposition result of (0,1,6), the corresponding sequential components would be converted into RDFFC. The port encoding in our PDK is reported in Table.V.

The marked results are presented in the Table for the above FSM example.VI and its corresponding mapped netlist are depicted in Fig.11.

Table V. Port Encoding

| Type | Encoding | Description |
|---|---|---|
| SFQ set | 0 | Stores an SFQ in the loop |
| SFQ clear | 1 | Clear the SFQ in the loop |
| Toggle State | 2 | Reverse the SFQ state |
| Clk Out | 3 | Non-Destructive Read Out |
| Clk Out & SFQ clear | 4 | Destructive Read Out |
| Clk Negative Out & SFQ clear | 5 | Destructive Read Negative Out, used in SFQ NOT. |
| Clk Out & Clk Negative Out & SFQ clear | 6 | Complementary Destructive Read Out, used in DFFC. |

Table VI. Marked Results

|  | Q1 | Q0 |
|---|---|---|
| Net | ($Q0 * En, En, Clk$) | ($Din, Rst, En$) |
| Type ID | (0,1,3) | (0,1,4) |
| Mapped gate | NDRO | RDFF |

Another example is when the complex circuit structure, such as RSFQ-AND (Fig.7), is found in the decomposition result. Then, the group of nodes will be converted into the corresponding gates. After gate mapping, the nets are also assigned in terms of the input connections, and the ports are assigned based on the information in the cell library. More details on this example are provided in Section IV-A. The overall algorithm for state decomposition and sub-FSM mapping is summarized in Algorithm 2.

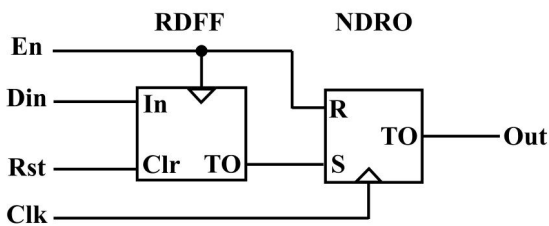

**Fig.11.** Mapped netlist of the example FSM (ERDFF).

**Algorithm 2:** State decomposition and sub-FSM mapping

1: **Input**: FSM, StatesTree, GateTable
2: **Output**: Sub-FSMs with type marks
3: **Function** decomposeAndMatch(FSM,StateEncoding)
4:   inTable←extractInputTransTable(FSM, StatesTree)
5:   outTable←extractOutputTable(FSM, StatesTree)
6:   optResultTable←initialize()
7:   netlist←initialize()
8:   **for** inputSig **in** inTable.keys **do**
9:     **for** currBit **in** States **do**
10:       expr←extractSoP(inputSig,inTable)
11:       optExpr←ROBDD(expr)
12:       optResultTable.update(currBit:optExpr)
13:   **for** outputSig **in** outTable.keys **do**
14:     **for** currBit **in** States **do**
15:       expr←extractSoP(outputSig,outTable)
16:       optExpr←ROBDD(expr)
17:       optResultTable.update(currBit:optExpr)
18:   **for** bit **in** optResultTable.keys **do**
19:     **for** expr **in** optResultTable[bit] **do**
20:       Mark the signal type based on expressions
21:   **for** bit **in** optResultTable.keys **do**
22:     Sort the markers
23:     mappedGate←GateTable[markers]
24:     **if** Failed **then**
25:       **return** failedFlag, expressions
26:     assign nets
27:     update to netlist
28:   **if** All succeeded **then**
29:     **return** succeedFlag, netlist
30: Create Multi-threading pool
31: resultList←initialize()
32: **for** stateEncoding **in** stateTree **do**
33:   task←decomposeAndMatch(FSM,StateEncoding)
34:   pool.add(task, result→resultList)
35: **if** succeedFlag **in** resultList **then**
36:   pool.terminate()
37:   **return** netlist

## G. Supergates Utilization

There are only two states for a 1-bit sequential component, and thus it is indecomposable by using the above method. If the decomposition result of a large FSM contains a sub-FSM, which cannot be mapped in the given Process Design Kit (PDK), we apply the supergate technique, in which the supergate is a small scale circuit that contains several elementary gates to build the target FSM. For example, the given state transition diagram (Fig.12) is not in our generic PDK. However, we build a supergate based on the state transition table and output equation using the existing cells in our PDK. This supergate netlist can be stored in the database inside the algorithm.

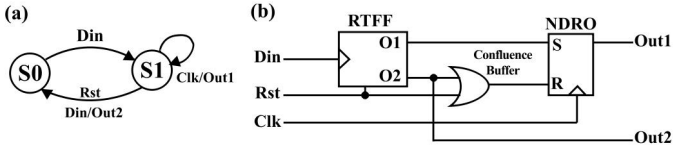

**Fig.12.** (a) State Transition Diagram and (b) its corresponding supergate, comprising RTFF, NDRO, and Confluence Buffer.

IV. EXPERIMENTS

*A. Small-Scale Case Studies*

To demonstrate how the synthesizer operates, we choose two small-scale circuits to display the synthesis process details, besides the ERDFF mentioned above.

(1) Feedback Loop Logic in Bit-Slice ALU

The details of the feedback loop logic are provided in [21]. After synthesis, the state transition diagram, and its best state encoding are illustrated in Fig.13, while the decomposition result is presented in Table VII. The synthesis result suggests that the FSM could be decomposed into three sub-FSMs, in which SEQ3, SEQ1, and the AND operation can be replaced as an RSFQ-AND gate (Fig.14). There is a feedback loop in the final netlist, feeding the output back to the input of SEQ3. The OR operation can be mapped with an RSFQ confluence buffer, similar to the OR gate in CMOS logic. Note that after technology mapping, only three RSFQ gates are used to compose this circuit.

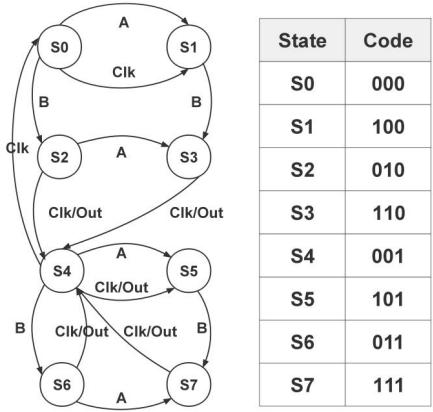

**Fig.13.** FSM of the feedback loop logic and its best state encoding derived from the synthesis result.

Table VII. Decomposition Results of Feedback Logic

|   | Q2 | Q1 | Q0 |
|---|---|---|---|
| A | $Q2^* = Q2$ | $Q1^* = Q1$ | $Q0^* = Q0 + A$ |
| B | $Q2^* = Q2$ | $Q1^* = Q1 + B$ | $Q0^* = Q0$ |
| Clk | $Q2^* = Q2 \cdot \overline{Clk} + Out$ | $Q1^* = Q1 \cdot \overline{Clk}$ | $Q0^* = Q0 \cdot \overline{Clk}$ |
| Out | | $Out = (Q2 \cdot Q0 + Q1) \cdot Clk$ | |

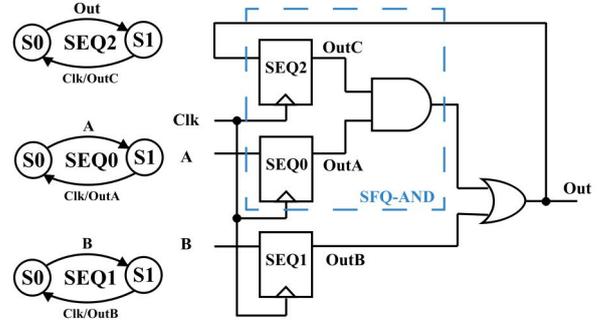

**Fig.14.** Sub-FSMs and the intermediate netlist. The SEQ3, SEQ1, and the AND operation will be replaced with an RSFQ-AND gate.

(2) 2-bit counter

The state transition diagram of the 2-bit counter and its best state encoding are depicted in Fig.15. The decomposition result is presented in Table.VIII.

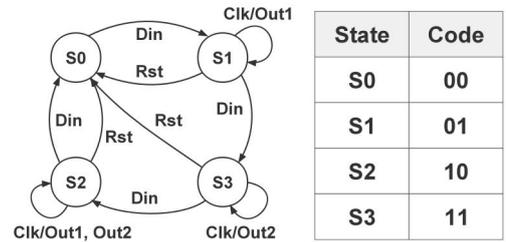

**Fig.15.** FSM of the 2-bit counter and its best state encoding derived from the synthesis result.

Table VIII. Decomposition Results of a 2-bit counter

|   | Q1 | Q0 |
|---|---|---|
| Din | $Q1^* = Q1 \oplus (Q0 \cdot Din)$ | $Q0^* = Q0 \oplus Din$ |
| Rst | $Q1^* = Q1 \cdot \overline{Rst}$ | $Q0^* = Q0 \cdot \overline{Rst}$ |
| Clk | $Q1^* = Q1$ | $Q0^* = Q0$ |
| Out | $Out1 = Q0 \cdot Clk$ | |
|     | $Out2 = Q1 \cdot Clk$ | |

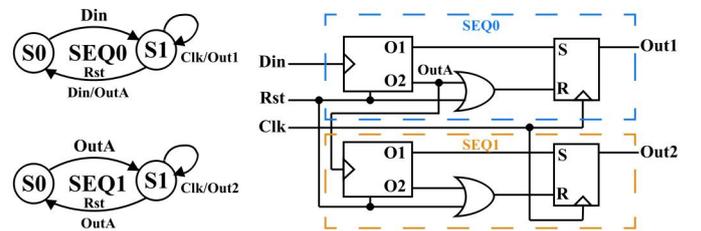

**Fig.16.** Sub-FSMs and the final netlist of the 2-bit counter. The supergate is applied in the mapping process.

After decomposition, there are two sub-FSMs that our PDK cannot map. Hence, the supergate technique is applied (Fig.12 in Section III-G), efficiently mapping both sub-FSM. The sub-FSMs and the netlist are presented in Fig.16. Another advantage of our decomposition algorithm is neglecting loops in the synthesis results of the counters, which means we can derive a serial counter regardless of its size. Also, since the post-synthesis clock frequency is mainly limited by the data path from the first RTFF to the final RTFF, the post-routing clock frequency can be improved by delaying the clock path from the first NDRO to the final NDRO. Thus, it is possible to derive a 32-bit counter with a post-routing clock frequency closer to a 2-bit counter.

*B. Benchmark Circuits Experiments*

To demonstrate the capability of synthesizing large-scale sequential circuits, we choose a 32-bit counter and the benchmark circuits from the ISCAS89 s-series. The results are summarized in Table IX. The Frequency the Power columns mean the post-synthesis clock frequency and the post-synthesis power dissipation, and the timing and power data of the gate used in the synthesis result from our PDK. Note that for the ISCAS s-series, we use the behavioral description [22] of these benchmark circuits to obtain the FSM. Since the circuit contains sequential and combinational blocks, only the former blocks are synthesized using the FSM decomposition algorithm. The combinational blocks are synthesized using the method presented in Section II-A.

The synthesis results highlight that using the FSM decomposition algorithm sharply reduces the number of gates of the counters and ISCAS s-series compared to [9]. This is because, in the traditional method, the Boolean gates build the combinational logic in the FSM, while in the RSFQ logic, the RSFQ gates are originally modeled using small-scale FSMs. So the synthesis becomes a two-step process: (1) Using the FSM of the RSFQ gates to emulate the Boolean gates and (2) Using the RSFQ Boolean gates to build the combinational logic in the target FSM. Since our algorithm cancels the intermediate step of the RSFQ gates to the Boolean logic and uses the small FSM of the RSFQ gates to compose the target FSM, theoretically, the gates' power consumption is lower.

Table IX. Summary of the Synthesis Results

| Circuits | No. of Junctions | No. of Gates | Frequency (GHz) | Power (μW) |
|---|---|---|---|---|
| ERDFF | 23 | 2 | 73.5 | 5.6 |
| Feedback logic | 19 | 3 | 41.6 | 4.1 |
| 2-bit counter | 52 | 6 | 128.5 | 11.9 |
| 4-bit counter | 104 | 12 | 42.8 | 23.9 |
| 32-bit counter | 864 | 96 | 4.1 | 190.2 |
| s208.1 | 338 | 33 | 18.4 | 72.8 |
| s298 | 803 | 85 | 45.2 | 161.2 |
| s344 | 1215 | 134 | 36.8 | 283.5 |

To verify the functionality of the synthesis result, we chose the 32-bit counter for the Verilog simulation, and the corresponding waveform results are depicted in Fig.17. This trial proves that our method's functionality is correct. The frequency of the input signal data is not limited except for the timing constraints, i.e., the setup time and hold time constraints. Moreover, the output result is sampled with the same frequency as the global clock because there are no loops in the synthesized netlist of this 32-bit counter.

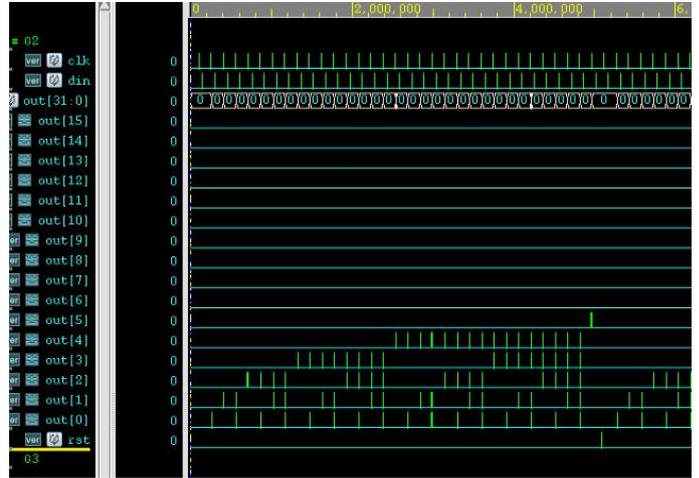

**Fig.17.** Waveform of a 32-bit counter.

V. CONCLUSION

This paper presents a novel algorithm for synthesizing sequential circuits for RSFQ logic. The developed algorithm employs the FSM decomposition method from the RSFQ perspective, accomplishing the sequential circuit synthesis using the FSM of the RSFQ gates to build a large FSM system. Besides, state encoding, sub-FSM mapping, and supergates utilization are also introduced to complete the algorithm. Our method significantly reduces the number of junctions by around 70% in each circuit compared with the traditional methods after synthesis, with the maximum frequency reaching up to the limitation of our PDK. Furthermore, the simulation results on a 32-bit counter verify our method's functionality. By employing the FSM decomposition method, the RSFQ sequential synthesis problem could be solved and the circuit performance including clock frequency, area and power consumption could be improved.

ACKNOWLEDGMENT

The authors would like to thank Ruoting Yang for providing the verilog models for simulation.

REFERENCES

[1] K. K. Likharev and V. K. Semenov, "RSFQ logic/memory family: A new Josephson-junction technology for sub-terahertz-clock-frequency digital systems," IEEE Trans. Appl. Supercond., vol. 1, no. 1, pp. 3–28, Mar. 1991.
[2] N. K. Katam and M. Pedram, "Logic Optimization, Complex Cell Design, and Retiming of Single Flux Quantum Circuits," in IEEE Transactions on Applied Superconductivity, vol. 28, no. 7, pp. 1-9, Oct. 2018, Art no. 1301409, doi: 10.1109/TASC.2018.2856833.

[3] G. Pasandi, A. Shafaei, and M. Pedram. "SFQmap: A Technology Mapping Tool for Single Flux Quantum Logic Circuits." International Symposium on Circuits and Systems (ISCAS) 2018.

[4] G. Pasandi, et al. "PBMap: A Path Balancing Technology Mapping Algorithm for Single Flux Quantum Logic Circuits." IEEE Transactions on Applied Superconductivity 29.4(2018):1-14.

[5] S. Yamashita, K. Tanaka, H. Takada, K. Obata and K. Takagi, "A transduction-based framework to synthesize RSFQ circuits," Asia and South Pacific Conference on Design Automation, 2006., 2006, pp. 7 pp.-, doi: 10.1109/ASPDAC.2006.1594693.

[6] D. Shah, E. Hung, C. Wolf, S. Bazanski, D. Gisselquist and M. Milanovic, "Yosys+nextpnr: An Open Source Framework from Verilog to Bitstream for Commercial FPGAs," 2019 IEEE 27th Annual International Symposium on Field-Programmable Custom Computing Machines (FCCM), 2019, pp. 1-4, doi: 10.1109/FCCM.2019.00010.

[7] B. L. Synthesis, "ABC: A system for sequential synthesis and verification," Berkeley Logic Synth. Verif. Group, 2011.

[8] A. Mishchenko, S. Chatterjee, R. Brayton, X. Wang, and T. Kam, "Technology mapping with boolean matching, supergates and choices," Technical Report, UC Berkeley, 2005.

[9] G. Pasandi and M. Pedram, "qSeq: Full Algorithmic and Tool Support for Synthesizing Sequential Circuits in Superconducting SFQ Technology," 2021 58th ACM/IEEE Design Automation Conference (DAC), 2021, pp. 133-138, doi: 10.1109/DAC18074.2021.9586102.

[10] S. V. Polonsky, V. K. Semenov and A. F. Kirichenko, "Single flux, quantum B flip-flop and its possible applications," in IEEE Transactions on Applied Superconductivity, vol. 4, no. 1, pp. 9-18, March 1994, doi: 10.1109/77.273059.

[11] X. Gao, et al. "Design and Verification of SFQ Cell Library for Superconducting LSI Digital Circuits." IEEE Transactions on Applied Superconductivity PP.99(2021):1-1.

[12] S. Takamaeda-Yamazaki, (2015). Pyverilog: A Python-Based Hardware Design Processing Toolkit for Verilog HDL. In: Sano, K., Soudris, D., Hübner, M., Diniz, P. (eds) Applied Reconfigurable Computing. ARC 2015. Lecture Notes in Computer Science(), vol 9040. Springer, Cham. https://doi.org/10.1007/978-3-319-16214-0_42

[13] Y. Shi, C. W. Ting, B. Gwee and Y. Ren, "A highly efficient method for extracting FSMs from flattened gate-level netlist," Proceedings of 2010 IEEE International Symposium on Circuits and Systems, 2010, pp. 2610-2613, doi: 10.1109/ISCAS.2010.5537093.

[14] R. Kibria, N. Farzana, F. Farahmandi and M. Tehranipoor, "FSMx: Finite State Machine Extraction from Flattened Netlist With Application to Security," in 2022 IEEE 40th VLSI Test Symposium (VTS), San Diego, CA, USA, 2021 pp. 1-7. doi: 10.1109/VTS52500.2021.9794151

[15] S. Devadas, and A. R. Newton . "Decomposition and factorization of sequential finite state machines." IEEE Transactions on Computer-Aided Design of Integrated Circuits and Systems 8.11(1989):1206-1217.

[16] P. Szotkowski, M. Rawski , and H. Selvaraj . "A Graph-Based Approach to Symbolic Functional Decomposition of Finite State Machines." International Conference on Systems Engineering IEEE, 2009.

[17] W. -. Yang, R. M. Owens and M. J. Irwin, "Multi-way FSM decomposition based on interconnect complexity," Proceedings of EURO-DAC 93 and EURO-VHDL 93- European Design Automation Conference, 1993, pp. 390-395, doi: 10.1109/EURDAC.1993.410666.

[18] O. Biggar, M. Zamani , and I. Shames . "Modular Decomposition of Hierarchical Finite State Machines. " (2021).

[19] M. P. Desai, H. Narayanan , and S. B. Patkar . "the realization of nite state machines by decomposition and the principal lattice of partitions of a submodular function." (2017).

[20] C. Drake, "PyEDA: Data Structures and Algorithms for Electronic Design Automation," in Proceedings of the 14th Python in Science Conference (SciPy 2015), 2015, pp. 26–31.

[21] G. Tang. "Studies on Datapath Circuits for Superconductor Bit-Slice Microprocessors." (2016).

[22] ISCAS High-Level Models [Online]. Avaliable: https://web.eecs.umich.edu/~jhayes/iscas.restore